\documentclass[pageno]{jpaper}

\usepackage[normalem]{ulem}
\usepackage{graphicx}
\usepackage{bbding}
\usepackage{multirow}
\usepackage[linesnumbered,boxed]{algorithm2e}
\usepackage{hyperref}
\usepackage{breakurl}
\usepackage{url}

\begin{document}

\title{
MIMS: Towards a Message Interface based Memory System}

\date{}

\author{Licheng Chen, Tianyue Lu, Yanan Wang, Mingyu Chen, Yuan Ruan, \\ Zehan Cui, Yongbing Huang, Mingyang Chen, Jiutian Zhang, Yungang Bao \\ State Key Laboratory of Computer Architecture, \\ Institute of Computing Technology, Chinese Academy of Sciences \\ \{chenlicheng,lutianyue,wangyanan,cmy,ruanyuan\}@ict.ac.cn \\ \{cuizehan,huangyongbing,chenmingyang,zhangjiutian,baoyg\}@ict.ac.cn}

%


\maketitle

\thispagestyle{empty}

\begin{abstract}
Memory system is often the main bottleneck in chip-multiprocessor (CMP) systems in terms of latency, bandwidth and efficiency, and recently additionally facing  capacity and power problems in an era of big data. A lot of research works have been done to address part of these problems, such as photonics technology for bandwidth, 3D stacking for capacity, and NVM for power as well as many micro-architecture level innovations. Many of them need a modification of current memory architecture, since the decades-old synchronous memory architecture (SDRAM) has become an obstacle to adopt those advances. However, to the best of our knowledge, none of them is able to provide a universal memory interface that is scalable enough to cover all these problems.

In this paper, we argue that a message-based interface should be adopted to replace the traditional bus-based interface in memory system. A novel message interface based memory system (MIMS) is proposed. The key innovation of MIMS is that processor and memory system communicate through a universal and flexible message interface. Each message packet could contain multiple memory requests or commands along with various semantic information. The memory system is more intelligent and active by equipping with a local buffer scheduler, which is responsible to process packet, schedule memory requests, and execute specific commands with the help of semantic information. The experimental results by simulator show that, with accurate granularity message, the MIMS would improve performance by 53.21\%, while reducing energy delay product (EDP) by 55.90\%, the effective bandwidth utilization is improving by 62.42\%. Further more, combining multiple requests in a packet would reduce link overhead and provide opportunity for address compression.

\end{abstract}

\section{Introduction}
The exponential growth of both the number of cores/threads (computing resources) and the amount of data (working set) demands high memory throughput and capacity. The number of cores integrated into one processor chip is expected to double every 18 months \cite{ITRSRoadmap}, many-core processors have been on the market, such as 60-core Intel Xeon Phi Coprocessor \cite{XeonPhi}, and  100-core Tilera TILE-GX 64-bit processor \cite{TileraGX}. The increasing number of cores would result in severe bandwidth pressure on the memory system. Memory requests from multiple cores would also interfere with each other and result in low locality. Thus future DRAM architectures place a lower priority on locality and a higher priority on parallelism \cite{RethinkDRAMArch}. On the other hand, the amount of data is predicted to grow with a rate of 40\% per year \cite{BigData}, it has become a hot topic in both academic and industry communities recent years. Big data processing requires more memory capacity and bandwidth.

However main memory that acts as the bridge between high level data and low level processor is failed to scale, leading memory system to be a main bottleneck. Besides the well-known memory wall problem \cite{MemoryWall}, the memory system also faces many other challenges (walls), which are concluded as followed:

\begin{bfseries}Memory wall (Latency)\end{bfseries}: The original "memory wall" referred to memory access latency problem \cite{MemoryWall} and it was the main problem in memory system until mid-2000s when the CPU frequency race slowed down. Then came the multi/many core age. The situation has changed a bit that queuing delays have become a major bottleneck, and might contribute more than 70\% of memory latency \cite{RethinkDRAMArch}. Thus for future memory architecture, it should place a higher priority to reduce queuing delays. Exploiting higher parallelism in memory system could reduce queuing delays because it is able to de-queue requests faster \cite{RethinkDRAMArch}.

\begin{bfseries}Bandwidth wall\end{bfseries}: The increasing number of concurrent memory requests along with the increasing amount of data, result in heavy bandwidth pressure. However the bandwidth of memory is failing to scale due to the relatively slow growth of pin counts of processor module (about 10\% per year \cite{ITRSRoadmap}). This has been concluded as bandwidth wall \cite{BandwidthWall}. The average memory bandwidth for each core is actually decreasing. In DDRx memory system, the memory controller (often integrated on processor chip) connects directly with DRAM devices through wide synchronous DDRx bus, each channel costs hundreds of processor pins.  Using narrower and higher speed serial bus between processor and memory devices could alleviate the pin count limitation, such as Full-buffered DIMM and its successors. Optical interconnection and 3D stacking are supposed to solve this problem substantially in the future.

\begin{bfseries}Efficiency problem\end{bfseries}: Latency and bandwidth are only physical factors of a memory system while the efficiency of memory access really counts for Processor. Current memory controller normally accesses a cache-block data (64B) in a BL burst (BL is 8 for DDR3) and DRAM modules activates a whole row (e.g. 8KB) in a bank. These large and fixed-size designs help to increase the peak bandwidth when memory access has a good spatial locality. However, in multiple core system, the locality both in row buffer and cache line is decreased \cite{RethinkDRAMArch,AGMS}. It has been shown that large data cache had almost no benefit for scale-out workload performance \cite{CloudSuite,ScaleOutProcessors}.
For data access without spatial locality, coarse-grained data unit would waste activate power and reduce effective bandwidth since it may move data that never be used. This is known as overfetch problem. Memory system that supports fine granularity access \cite{AGMS,DGMS} could improve bandwidth efficiency.



\begin{bfseries}Capacity wall\end{bfseries}: Big data requires higher memory capacity. But the slowly growing of pin count limited the number of memory channels each processor could support. Furthermore the number of DIMM (dual in-line memory module) could be supported in each channel is limited due to signal integrity restriction. For instance, only one DIMM is allowed in DDR3-1600 while four DIMMs are allowed for DDR1\cite{DIMMLimit}. And the capacity each DIMM could provide is growing slowly due to the difficulties in decreasing the size of a DRAM cell's capacitors\cite{BOBMemory}. To increase the DIMM counts in single channel, registers and buffers are added to the memory interface{RDIMM,LRDIMM}.
There are also various proposals to provide big memories, such as BOB (Buffer On Board) Memory \cite{BOBMemory}, HMC (Hybrid Memory Cube) \cite{HMC}, high density non-volatile memory (e.g. PCM) \cite{PCMArch} and 3D-Stacked memory \cite{PacketMemory}.


\begin{bfseries}Power wall\end{bfseries}: memory system has been reported to be the main power consumer in servers, contributing about 40\% to the total system power \cite{MemoryPowerWall,MemScalePower}. Capacitor-based memory cell contributes the main power to the DRAM system which is not an architectural issue. However, due to the overfetch problem, a large part of the dynamic power of DRAM is wasted. Improving memory system to support Sub-Access in row buffer and fine granularity memory access could alleviate the overferch problem. To reduce static power, non-volatile memory (e.g. PCM) could be investigated as potential alternatives for existing memory technologies. NV-memory  has a totally different access parameters, so it can not work under a memory interface designed for DRAM such as DDRx.

Besides all the above walls, there is a long trend to equip the memory system with some simple processing logic to make memory more autonomous. Logic in memory would significantly reduce data movement between processor and memory system, improve performance and decrease power consumption. Many autonomous memory systems have been proposed before, such as Processing in Memory (PIM) \cite{PIMArch}, Active Memory Operation (AMO) \cite{AMOperations,AMController}, and Smart Memory \cite{HuaweiSmartMemory}. These technologies are limited to proprietary designs and never get into a standard memory interface with read and write operations only.

In summary, a lot of works have been done to alleviate various memory system bottlenecks. However each of them is only focus on one or part of these walls. To the best of our knowledge, none of them is able to address all these problems. Table 1 lists different approaches and which problem each addressed (please refer to section 2 for detail). For example, the BOB \cite{BOBMemory} memory system is focus on address bandwidth and capacity wall, and the simpler controller is responsible to schedule requests, makes the memory system a little autonomous.

\begin{table}[h!]
  \centering
  \begin{tabular}{|c|c|c|c|c|c|c|}
    \hline
    \textbf{} & \textbf{LY} & \textbf{BW} & \textbf{CY} & \textbf{EY} & \textbf{PR} & \textbf{AS}\\
    \hline
    \hline
    Sub-Access & * & $\times$ & $\times$ & * & ${\surd}$ & ${\times}$\\
    \hline
    FGMS & * & $\times$ & $\times$ & ${\surd}$ & ${\surd}$ & $\times$\\
    \hline
    Buffer-Chip & $\times$ & $\times$ & ${\surd}$ & $\times$ & $\times$ & $\times$\\
    \hline
    BOB MS & $\times$ & ${\surd}$ & ${\surd}$ & $\times$ & $\times$ & *\\
    \hline
    AMO/Smart & $\times$ & $\times$ & $\times$ & $\times$ & ${\surd}$ & ${\surd}$\\
    \hline
    NVM & $\times$ & $\times$ & ${\surd}$ & $\times$ & ${\surd}$ & *\\
    \hline
    3D-Stacked & $\times$ & * & ${\surd}$ & $\times$ & ${\surd}$ & *\\
    \hline
    Photonics & $\times$ & ${\surd}$ & ${\surd}$ & * & $\times$ & *\\
    \hline
    Mobile-DRAM & $\times$ & $\times$ & * & $\times$ & ${\surd}$ & *\\
    \hline
    \hline
    \multicolumn{7}{|c|}{Note: ${\surd}$-Yes, $\times$-No, *-Maybe.} \\
    \multicolumn{7}{|c|}{ LY: Latency, BW: Bandwidth, EY: Efficiency,} \\
    \multicolumn{7}{|c|}{CY: Capacity, PR: Power, AS: Autonomous} \\
    \hline
  \end{tabular}
%
  \caption{Comparison of different approaches to alleviate different walls. }
  \label{table:formatting}
\end{table}

In this work, we argue that traditional synchronous bus-based memory interface should be redesigned to incorporate future innovations. In contrast to traditional read or write bus transaction based memory interface, a flexible asynchronous message based memory interface will bring more design opportunity. We propose a uniform message interface based memory system (MIMS): Memory request and response are sending via asynchronous messages. A local buffer scheduler is put between memory controller and memory devices. Device-specific scheduling is decoupled from CPU memory controller, which is only responsible to compose memory requests for packet. The memory controller communicates with buffer scheduler over high-speed serial point-to-point link with a flexible message packet protocol. Each message could contain multiple memory requests or responses as well as semantic information such as granularity, thread id, priority, timeout. The buffer scheduler act as the traditional memory controller: it needs to track status of local memory devices, schedule requests, generate and issue specific DDR commands, meanwhile meeting the timing constraints (such as DDR3 for DRAM devices). Additionally, buffer scheduler can use various semantic information from the CPU to help its scheduling.



MIMS will bring at least the following advantages:

\begin{enumerate}
\item It provides a uniform, scalable message interface to access different memory system. The status tracking and request scheduling is decoupling from memory controller and pushed down to buffer scheduler. Thus the integrated memory controller has no timing limitations and could easily scale to other emerging memory technologies. Also the memory capacity is only restricted by the buffer scheduler, which is decoupled with memory controller.
\item It could naturally support variable granularity memory requests. Each memory request is transferred with  the exact size of really useful data. This could significantly improve data/bandwidth effectiveness and reduce memory power consumption.
\item It enables inter-requests optimization during the memory access such as combining multiple operations in a packet and compressing memory address for a sequence of requests.
\item It is easy to add additional semantic information to the message to help the buffer scheduler to make decisions when doing local scheduling. Local computation or intelligent memory operation requests can also be added as part of the message.
\end{enumerate}

To demonstrate the benefits of using a message interface, we have implemented a cycle-detailed memory system simulator, MIMSim. Experiments on fine-granularity access, trunk memory request and address compression are taken. The results provide elementary proof for the benefits of MIMS.

The rest of the paper is organized as follows. Section 2 gives background on memory system and related work, while Section 3 presents the Message Interface based Memory System, includes architecture, packet format, address compression and challenge. Section 4 presents the experimental setup, and the results are presented in section 5. Section 6 gives a conclusion of this paper.


\section{Background and Related Work}

In this section, we first give a brief description of the most commonly used JEDEC DDRx SDRAM memory systems, and then discuss some optimizations on memory architecture, including Sub-Access, buffer-chip memory, autonomous memory, and some aggressive memory system, such as non-volatile memory (NVM), 3D-stacked memory, photonics interconnect memory and mobile-DRAM in server.

\subsection {DDRx Memory System}

The JEDEC standardized DDR (Double Data-Rate) \cite{DDR3Specification} synchronous DRAM is dominated nowadays. In DDRx memory system, memory controller could be considered as the bridge logic between processor and DRAM devices, which is responsible to receive memory requests from last level cache (LLC). The memory controller needs to track the status of DRAM devices (e.g., bank states) and generates DRAM commands for each selected requests meanwhile meeting the DDRx timing constraints. The integrated memory controller communicates directly with DRAM devices over wide synchronous DDR bus with separate data, command, and address bus. This directly-connected design would result in high processor pin-count cost, and this has become a main bottleneck to support large memory capacity, because the growth of processor pin-count is failed to keep up with demand.

The memory system has a hierarchical organization, with different level parallelism. Each memory controller might support multiple memory channels, while each channel has separate DDR bus. Thus each channel could be accessed independently. Within a memory channel, there might be multiple DIMMs (Dual Inline Memory Module). Each DIMM might comprise with multiple ranks (1-4), and each rank provides a logical 64-bit data-path (bus) to memory controller (72-bit in ECC-DIMM). Multiple DRAM devices within a rank need to be operated in tandem. The x8 DRAM devices are commonly used today and they will be used in this work by default.

A prevalent DRAM device (chip) consists of multiple DRAM banks (8 in DDR3) which can be processed concurrently. There is a two dimensional array, consisting of rows and columns, within each DRAM bank. A row buffer is dedicated to each bank, which is usual 4-16KB. Before each column access, a row needs to be loaded into the row buffer by an active command. If latter requests are hit in the row buffer, it could be accessed directly by read/write command. Otherwise, the row buffer needed firstly to be precharged back to array before issuing a new request.

\subsection {Sub-Access Memory}

Sub-Access memory refers to dividing a whole component into multiple sub-components, so that each memory request only needs to access a portion of data. Here component could be rank, row buffer and cache line (FGMS).

Sub rank memory divides a memory rank into multiple logical sub-ranks, which could be accessed independently. Data layout might be adjusted so that each cache line is put in a sub-rank. Each memory access would only require a part of memory devices (in the same sub-rank) to be involved. Many different approaches could be classified as sub-rank, including Rambus's Module threading \cite{ThreadedMemoryModule}, Multicore DIMM (MCDIMM) \cite{MulticoreDIMM,SubRank}, and mini-rank \cite{MiniRank,HeteMimiRank}, heterogeneous Multi-Channel \cite{HeneMC}. Sub-rank technology could save memory power by alleviating the over-fetch problem and improve memory level parallelism (MLP). The downside of it is that the memory access latency will increase  since
part of the total device bandwidth can be utilized, because they still use coarse granularity memory access.

Udipi et.al \cite{RethinkDRAMArch} proposed SSA to reduce power. An entire cache line is fetched from a single subarray by re-organizing the data layout in SSA. Cooper-Balis and Jacob \cite{FineGrainedActive} proposed a fine-grained activation approach to reduce memory power, which only actives a smaller portion of row within the data array by utilizing posted-CAS command.

\begin{bfseries}FGMS\end{bfseries}: AGMS and DGMS \cite{AGMS,DGMS} adopted sub-rank memory system to allow dynamically use fine or coarse granularity memory access. They also proposed some smart data layouts to support high reliability with low overhead. Convey-designed Scatter-Gather DIMMs \cite{ConveySGDIMM} are promising to allow 8 bytes (fine granularity) access that could reduce the inefficiencies in nonunity strides or randomly memory accesses, however the implementation detail of SGDIMM is lack. Cray Black Widow \cite{CrayBlackWidow} adopted many 32-bit wide channels, allowing it to support 16B minimum access granularity.

\subsection {Buffer-Chip Memory}

To alleviate memory capacity problem, a common way is to put an intermediate buffer (logic) between the memory controller and DRAM devices, which could reduce the electrical load on the memory controller and improve signal integrity.

In Registered DIMM (RDIMM) \cite{RDIMM}, a simple register is integrated on DIMM to buffer control and address signals. Load Reduced DIMM (LRDIMM) \cite{LRDIMM} further buffers all signals that go to DRAM devices, including all data and strobes. Decoupled DIMM \cite{DecoupledDIMM} adopts a synchronization buffer to convert between low speed memory devices and high data rate memory bus. With a similar idea, BOOM \cite{BOOM} adds a buffer chip between the fast DDR3 memory bus and wide internal bus, which enable the use of low-frequency mobile DRAM devices, thus BOOM could save memory.

In Fully-Buffered DIMM (FBDIMM) \cite{FBDIMM_Arch,FBDIMM_Datasheet} memory module, there is an AMB (Advanced Memory Buffer) integrated on each DIMM module, multiple FBDIMMs are organized as a daisy chain which could support high capacity. The memory controller communicates with AMB through point-to-point narrow, high-speed channels with some simple packet protocol. Intel Scalable Memory Interface (SMI) \cite{SMBDatasheet} and IBM power 7 memory system \cite{IBMPower7,IBMAMB} also place logic buffers between the DRAM and the processor, which could support more memory channels.

Cooper-Balis et. al \cite{BOBMemory} proposed a generalized Buffer On Board (BOB) memory system. In BOB, intermediate logic is placed (on motherboard) between on-chip memory controller and DRAM devices. The memory controller communicates with the intermediate buffer through serial link. The memory controller is decoupling from scheduling, and the intermediate buffer actually acts as a traditional memory controller: it tracks status of its local memory devices and schedules memory requests, issues corresponding DRAM commands meanwhile meets timing constraints. The BOB memory system is promising to alleviate the capacity and bandwidth problem.

UniMA \cite{UniMA} aims to enable universal interoperability between processors and memory modules. Each memory module was equipped with a Unified DIMM Interface Chip (UDIC). The memory controller sends read/write requests to UDIC through the unified interface without worrying about any devices status or timing constraints.

\subsection	{Autonomous Memory}

There has been a long time effort to make memory autonomous by equipping main memory with processing logic to support some local computations. Processor-in-memory (PIM) systems incorporate processing units on modified DRAM chips \cite{PIMArch,PIMDataArch}. Active memory controller \cite{AMController,AMOperations} adds an active memory unit to support active memory operation (AMO), such as scalar operations (e.g. inc, dec) and stream operations (e.g. sum, max). Smart Memory \cite{HuaweiSmartMemory} attaches simple compute and lock with data, thus reduces chip I/O bandwidth and achieves high performance and low latency.

\subsection	{Aggressive Memory System}

Non-volatile memory (such as PCM) \cite{PCMArch,HybridPCMIBM,PCMMemoryPitt,Hybrid3DPCM} has been considered as a potential replacement for DRAM chip in the future. NVM devices could remove static power consumption and promise to provide higher capacity. Recent result shows that some NVM has comparable latency to the DRAM. However NVM chip usually have a different timing requirement so that it can not be incorporated into the DDRx memory interface directly.

On Chip optical interconnection is expected to provide enough bandwidth for memory system. Udipi et al. \cite{PacketMemory} proposed a novel memory architecture uses photonics interconnects among memory controller with 3D memory stacked dies, they also proposed a novel packet based interface to relinquish the memory controller and allow the memory modules to be more autonomous. Each packet only contains one memory requests and is processed in a FCFS order.

Hybrid Memory Cube (HMC) \cite{HMC} Utilizes 3D interconnect technology. A small logic layer that sits below vertical stacks of DRAM die connected by through-silicon via (TSV) bonds. This logic layer is responsible to control memory devices. The memory controller communicate to HMC logic chip via abstracted high-speed interface, logic layer flexibility allows HMC cubes to be designed for multiple platforms and applications without changing the high-volume DRAM.

\section {Message Interface based Memory System}

\subsection {Why use a Message-based Interface?}

The current bus-based memory interface can be dated back to the 1970s when the first DRAM chip in the world was produced. After 40 years the main characteristics remains unchanged: separated data, address and control signals; fixed transfer size and memory access timing (latency); CPU being aware and take care of every bit of storage on the memory chip; limited on-going operations on the interface. One may argue that a simple and raw interface for DRAM keeps the minimum latency for memory access, but it also brings obstacles to improve memory performance as described in Introduction section. Nowadays with more and more parallelism in computer system, single memory access latency is not the main issue for overall performance any more. Is it the right time to change this decades-old interface?

Decoupling is the common trend of many previous works mentioned in section 2. That is to separate the data transfer and data organization. The CPU should only take care of sending requests and receiving data while a buffer controller takes charge of scheduling and local DRAM chip organization. A packet-based interface will enable this separation by encapsulating data, address and control signals.
If we just stop here, then packeting is only a low-level encapsulation for bus transactions.

We can go a further step from packet to message. Here message means the content of a packet is not predefined or fixed but programmable. Message also means CPU can put more semantic information on a packet other than simple read/write operations. Then the buffer controller can make use of this information to get better performance. The information maybe size, sequence, priority, process id, etc. or even array, link and lock. It is like that the buffer controller is integrated with CPU virtually to get all those necessary information for memory optimization.
A message-based interface will provide many opportunities to help solve memory system issues.

For latency problem, though a message interface may increase the latency of single operation, it helps to increase parallelism and do better prefetching and scheduling with semantic information, so contributes to decrease the overall latency.

For bandwidth problem, message interface will support better memory parallelism to full utilize the bandwidth; packet interface enables new interconnection technologies. Message also enables effective compression.

For efficiency problem, exact data size information will help reduce waste of over-fetch; message also enables exact prefetching to reduce unnecessary operations.

For capacity problem, decoupling enables special design for large capacity
memory systems; message even enables a network of extended memory systems.

For power problem, message enables fine-grained control of active DRAM regions. Decoupling also enables low power NVRAM to be included in memory system transparently.

For autonomous operation, message provides a natural support with semantic information.

To demonstrate the benefits of message interface, a draft architecture design  and  evaluation are given as following. It should be noted that the design and evaluation are elementary and incomplete to cover all the advantage of message -based interface.

\subsection {MIMS Architecture}

\begin{figure*}
  \centering
  \includegraphics[width=0.8\textwidth]{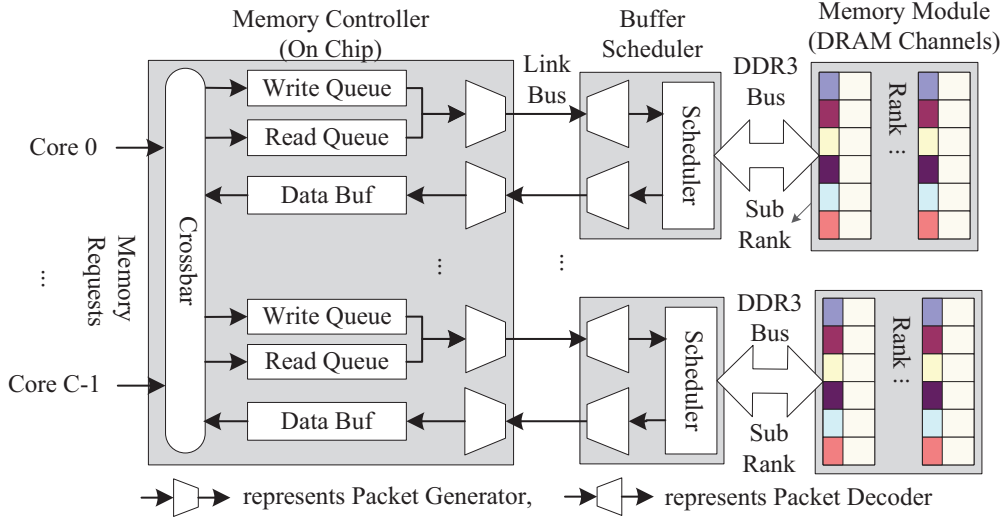}
  \caption{The Message Interface Memory System architecture.}
  \label{MIMS_Arch}
\end{figure*}

Figure \ref{MIMS_Arch} shows the architecture of the Message Interface based  Memory System (MIMS). As in the Buffer-On-Board (BOB) memory system \cite{BOBMemory}, the memory controller in processor does not directly communicate with memory devices (DRAM), instead, it communicates with the buffer scheduler via serialized point-to-point link which is narrower and could work at a much higher frequency. Each memory controller could support multiple buffer schedulers. Each buffer scheduler consists of memory request buffer, packet generator, packet decoder, return buffer and link bus.

The memory controller receives variable-granularity memory requests from multiple processor cores. The memory controller firstly chooses the target buffer scheduler based on the address mapping scheme, and then put it into on-chip network. The NOC routes each memory request into its target request buffer. For each buffer scheduler, the request buffers are divided into Read Queue and Write Queue, which is used to buffer read and write requests respectively. Read requests have high priority when scheduling requests to pack until the number of write requests in the write queue exceeds the high water mark \cite{VirtualWriteQueue,StageReads}. Then write requests get high priority, and write requests would be contiguously selected to be packed and sent to the corresponding buffer scheduler until the number of write requests return below the low water mark.

The Packet Generator is responsible to select multiple memory requests to put into a packet, send it to SerDes Buffer, construct the packet head, which containing meta data of a packet. Note that, the packing operation is not in the critical path, because the Packet Generator keeps tracking the status of the serialized link bus, and could start packing process in advance before the link bus become available (free). After the packet has been constructed and the link bus become available, the packet would be sent to the target buffer scheduler.


After receiving a message packet, the packet decoder on the buffer scheduler would unpack the packet and retrieve all memory requests integrated in the packet, and then send them to the scheduler. The scheduler acts as a traditional memory controller: it communicates with the DRAM memory module through wide and relative low frequency bus with the synchronous DDR3 protocol as in the traditional memory system. The scheduler needs to track all the memory module states (e.g. bank state, bus state) attached to it, schedules the memory requests and generates DRAM commands (such as ACTIVE, PRECHARGE, READ, WRITE etc.) based on the memory states, and issues the DRAM commands to the memory module meanwhile fulfills the DDR3 protocol constraints.

Sub-rank memory system is used to support variable-granularity memory access, which is similar as DGMS \cite{DGMS}. We use x8 DRAM devices, and each rank is separated into 8 sub-ranks, with one device per sub-rank. The data burst length (BL) is 8 in DDR3, thus the minimum granularity is 8B.


\subsubsection {Packet Format}

Message packet is the essential and critical component in MIMS, and packet should be designed to easily scale to support various memory optimizations. Each packet could support multiple memory requests. Each packet contains some Link Overhead (LKOH) which is generated and processed at lower layer such as link layer and physical layer. The LKOH is necessary for serial bus communication protocol, which usually contains reliability-related data such as Start, End signal, Sequence ID, and checksum codes (CRC).

The overhead of LKOH would be high if each packet only contains a small amount of data (payload). Especially for a read request, which only contains address and operation, the overhead of the LKOH would be near 50\% for a size of 8B LKOH (as in PCIe protocol). Combining multiple memory requests into a packet would increase the payload size.

\begin{figure}
  \centering
  \includegraphics[width=0.5\textwidth]{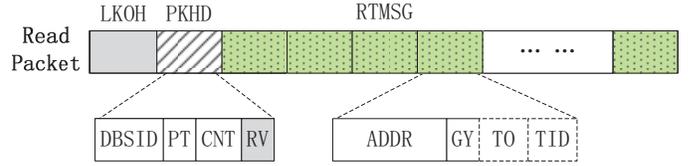}
  \caption{Read packet format.}
  \label{ReadPacket}
\end{figure}

To support multiple memory requests in a packet, we propose a variable-length packet format for MIMS. The packet has three basic types: Read Packet, Write Packet and Read-Return Packet, which might contain multiple read memory requests, write requests and return data respectively. Since each packet might contain variable number of requests, each packet is added a packet head (PKHD) which contains Meta data of the packet, the detail format of packet is shown in figure 4. We can see in figure \ref{ReadPacket} that a Read Packet has a packet head and multiple Request Messages (RTMSG), and it should contain LKOH. The packet head contains Destination Buffer Scheduler Identifier (DESID), Packet Type (PT, such as Read), the Count (CNT) of requests and some other Reserved (RSV) fields. Note that all the requests in a packet are sent to the same Destination Buffer Scheduler. After packet head, multiple request messages are closely aligned, each request message (RTMSG) represents a memory request with some other semantic message, it basically contains address (ADDR), granularity (GY) for each memory request, and could salable to contain more semantic message such as request timeout (TO), thread id (TID). The timeout require the longest acceptable latency (queue delay) that it must be scheduled and return, this is valuable to implement QoS for requests, other message that is valuable for scheduling could also be integrated in the RTMSG. All the RTMSGs in a packet are in the same format and same length, which makes to encode and decode reading packet  easily and effectively.

\begin{figure}
  \centering
  \includegraphics[width=0.5\textwidth]{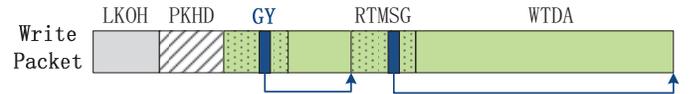}
  \caption{Write packet format.}
  \label{WritePacket}
\end{figure}

In a write packet, as shown in figure \ref{WritePacket}, the format is nearly in the same, it also contains LKOH, a packet head and multiple write requests, where the packet head is just the same as in the read packet, except that the packet type (PT) is Write Packet. Besides a RTMSG, each write request need also contains write data (WTDA). The RTMSG is the same with read request. Write data might be variable-length, and the length is determined by the granularity (in RTMSG) of the request. For example, the length of data is 8B for a fine granularity write, and it is 64B for a coarse granularity write.

Read-return packet has the same format with write packet. The request address needs to be returned since memory requests are scheduled out-of-order both in packet encoding and in buffer scheduler. In order to reduce the overhead of returning address, each read request could be assigned a request id, which is much smaller (10 bits is enough 1024 requests).

\subsubsection {Packet Decoding}

After the buffer scheduler received a packet, the Packet Decoder firstly reads the packet head and gets the meta data of the packet, such as DESID (Destination Buffer Scheduler ID), Packet Type, memory requests count and other reserved data. The DESID is used to check whether the packet was routed correctly. Then  the type of the packet and the count of memory requests are checked.

\begin{figure}
  \centering
  \includegraphics[width=0.45\textwidth]{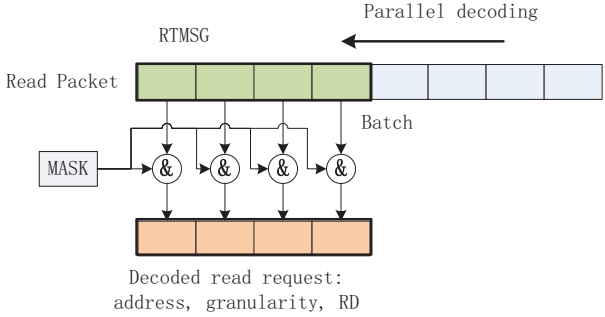}
  \caption{Parallel decoding for Read Packet. Each batch (4 in the figure) of RTMSGs are decoded in parallel, each RTMG could be decoded with a simple MASK operator.}
  \label{DecodeReadPacket}
\end{figure}

\begin{bfseries}Read packet\end{bfseries}: Since each read request messages has fixed-length fields, including address, granularity etc., it could be processed in parallel easily. Figure \ref{DecodeReadPacket} shows the process of parallel decoding for read packet. In this example, 4 RTMSGs are decoded once a time. Since the format for each RTMSG is same, they could be decoded with a single mask,  the address, granularity and other message of each read request could be extracted. After that, the next batch RTMSGs are ready to be decoded.

\begin{bfseries}Write Packet\end{bfseries}: Each write request has the Request Message along with variable length of data, where the length could be determined by the granularity of the memory request. For example, if the granularity is 4, then the length of data is 32B (8B * 4). The decoder process the requests in serial due to variable size: it extracts address, granularity and other message of the first write request, then it calculates the length of data base on the granularity, retrieves the write data, advances to the next request, until all the write requests are retrieved.


\subsubsection {Address Compression in a message packet}

Putting multiple memory requests in a packet provides a good opportunity to  address compression. It is also enables to compress data in write packet and  return packet. A lot of work had contributed to data compression. Motivated by that address would contribute about 39\% of packet (section 5.1 for more detail), we focus on compressing multiple address in a read packet which has not been investigated before, and we will show that with some simple compression algorithms, the address could be compressed efficiently.



\begin{figure*}
  \centering
  \includegraphics[width=0.8\textwidth]{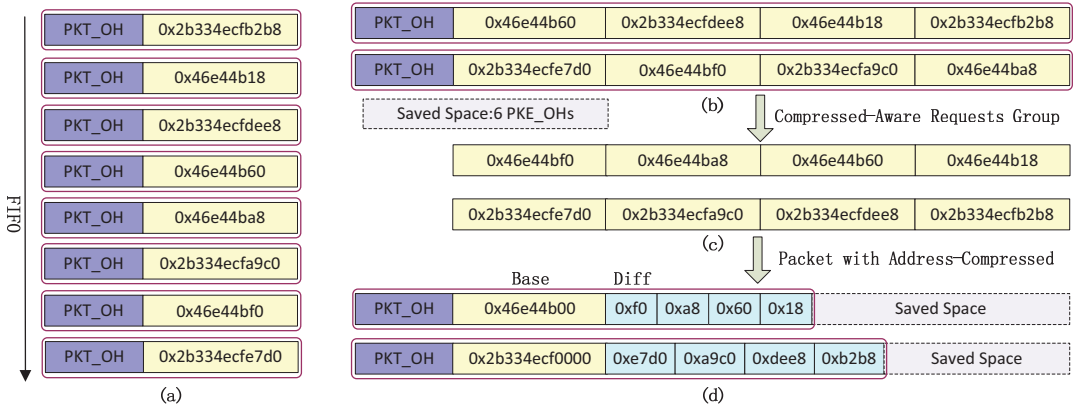}
  \caption{Conceptual example showing the benefit of involving multiple requests in a packet: packet overhead reduction and address compression. (a) FIFO with one request in each packet. (b) FIFO with multiple requests in each packet. (c) Out-of-order compressed-aware requests grouping. (d) Address compression in each packet.}
  \label{AddressCompression}
\end{figure*}

The example in figure \ref{AddressCompression} illustrates how involving multiple requests in a packet could reduce packet overhead and how address compression could further reduce the size of payload, thus reduce the demand to the bandwidth of link bus.Figure \ref{AddressCompression}(a) shows the simplified FIFO one request per packet, we can see that for 8 memory requests, it totally induces 8 packet overhead (PKT\_OH). And figure \ref{AddressCompression}(b) shows if the packet support to involve multiple requests, such as 4 requests in each packet, then there are 2 packets with induced 2 packet overheads, it could save 6 PKT\_OHs space. However since it still packet requests in FIFO, the addresses in each packet has relatively poor locality, which is obstacle to perform address compression. Thus in figure \ref{AddressCompression}(c), we selects memory requests in en-packet in an out-of-order and compress-aware manner, which firstly re-order memory requests and group multiple adjacent requests which is preferred to be selected in the same packet. Finally, figure \ref{AddressCompression}(d) shows the base-delta address compression in each packet, we choose a base, and all the address are then represented as the difference (DIFF) to the base, where the DIFF could have variable length, such as 2B in the first packet and 1B in the second packet.

%
%
%
%

\subsection {The challenge of designing a message interface based memory system}

Message based interface for memory system will bring challenges to all system levels that concerned with memory. Many challenges remain to be solved. Here is an incomplete list.
\begin{enumerate}
\item Complexity: Message processing is more complex than simple packet. Both the memory controller and buffer scheduler need more complex logic to accomplish the task, e.g. longer queue management and consistence checking. Although logic is becoming cheaper, it still needs to investigate whether the cost, power consumption, and increased latency can be controlled within an acceptable level.
\item ISA extension:  To full utilizing the flexibility of message, CPU needs to provide more semantic information along with read/write request. This may bring extensions need to the ISA. For example, how to provide the size information for variable granularity memory requests; how to deliver process information such as thread id, priority, timeout and prefetch; how to generate Active Memory Operation request to the memory controller.
\item Cache Support: To better support variable granularity memory accesses, variable-sized cache line is preferred though with difficulty. Sector cache for fine granularity and SPM (Scratchpad memory) for large granularity can also be used with a redesign.
\item  Programming: The semantic information may also be discovered and generated by software and sent via message. Application can be implemented with some hint API, or with the help of an aggressive compiler to generate MIMS special instructions automatically.
\end{enumerate}
%

\section{Experiments Setup}

\subsection{Simulator and Workloads}

To evaluate MIMS, we have implemented a cycle-detailed Message Interface Memory System simulator which is named MIMSim. We adopted DRAM modules (devices) based on DRAMSim2 \cite{DRAMSim2}, which is a cycle accurate DDR2/3 memory system simulator. The DRAMSim2 models all aspects of the memory controller and DRAM devices, including transaction queue, command queue and read-return queue, address mapping scheme, DDR data/address/command bus contention, DRAM device power and timing, and row buffer management. We add re-order buffer (ROB) to make simulation more accurate, the DRAM module is modified to support sub-rank. Channel interleaving address mapping is adopted as the default (baseline) configuration to maximum MLP (Memory Level Parallelism), and FRFCFS \cite{FRFCFSScheduling} scheduling policy with closed-page row buffer management.

\begin{figure}
  \centering
  \includegraphics[width=0.5\textwidth, height=2in]{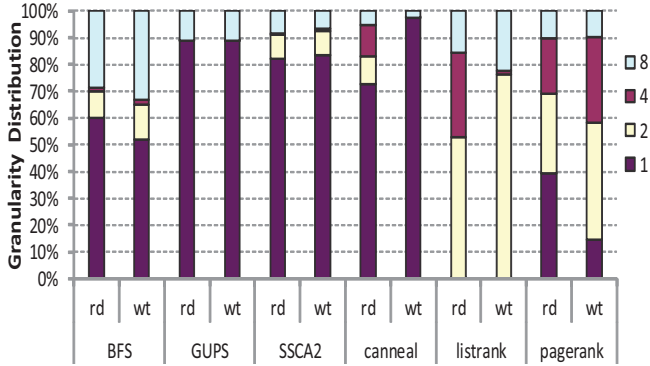}
  \caption{The read and write granularity distribution of FINE and MID memory-intensive workloads.}
  \label{GranularityDistribution}
\end{figure}

Pin \cite{Pin} is used to collect memory access traces from various of workloads running with 2-16 threads. We choose several multi-thread memory intensive applications from BFS in Graph500 \cite{Graph500}, PARSEC \cite{PARSEC}, Listrank \cite{ListRank}, Pagerank \cite{PageRank}, SSCA2 \cite{SSCA2}, GUPS \cite{GUPS}, NAS \cite{NAS}, STREAM \cite{STREAM}. Table \ref{Workloads} lists the main characteristics of these workloads. We classify the workloads into three categories based on the access granularity: fine granularity (FINE: <=3), Middle granularity (MID: 3-6), and coarse granularity (COR: 6-8). Memory read  and write requests are reported separately, including the read memory requests per kilo instruction (RPKI), the average read granularity (RG), the write memory requests per kilo instruction (WPKI), the average write granularity (WG), and the read/write ratio (RD/WT). The reason to separate the read and write characteristics is that we find the granularity distribution of read and write might different for some FINE and MID benchmarks. Figure \ref{GranularityDistribution} shows their granularity distribution. For example, in the canneal benchmark, the rate of 1-granularity is about 72.85\% for read requests, but it is about 97.59\% for write requests. And in the listrank benchmark, the rate of 2-granularity and 4-granularity is about 52.99\% and 31.54\% respectively for read requests, but they are about 76.51\% and 0.90\% respectively for write requests.

\begin{table}[h!]
  \centering
  \begin{tabular}{|c|c|c|c|c|c|c|}
    \hline
    \textbf{Cate.} & \textbf{Bench.} & \textbf{RPKI} & \textbf{RG} & \textbf{WPKI} & \textbf{WG} & \textbf{R/T} \\
    \hline
    \textbf{FINE} & GUPS & 69.67 & 1.78 & 69.62 & 1.78 & 1.00 \\
    \textbf{FINE} & SSCA2 & 20.89 & 1.68 & 20.42 & 1.56 & 1.02 \\
    \textbf{FINE} & canl. & 17.79 & 1.64 & 8.64 & 1.10 & 2.06 \\
    \textbf{FINE} & park. & 9.76 & 2.42 & 6.14 & 2.74 & 1.59 \\
    \hline
    \textbf{MID} & lirk. & 22.56 & 3.56 & 15.45 & 3.37 & 1.46 \\
    \textbf{MID} & BFS & 22.36 & 3.10 & 2.44 & 3.49 & 9.16 \\
    \hline
    \textbf{COR} & STRM. & 33.33 & 8.00 & 16.63 & 8.00 & 2.00 \\
    \textbf{COR} & bt & 7.68 & 7.98 & 7.63 & 7.98 & 1.01 \\
    \textbf{COR} & ft & 31.85 & 8.00 & 31.72 & 8.00 & 1.00 \\
    \textbf{COR} & sp & 8.04 & 7.98 & 7.89 & 7.98 & 1.02 \\
    \textbf{COR} & ua & 4.42 & 7.19 & 3.80 & 7.92 & 1.16 \\
    \textbf{COR} & ScPC. & 11.00 & 5.65 & 3.85 & 5.74 & 2.86 \\
    \textbf{COR} & perM & 2.62 & 6.28 & 2.40 & 6.12 & 1.09 \\
    \hline
    \hline
    \multicolumn{7}{|c|}{Note, canl.: canneal, park.: pagerank,} \\
    \multicolumn{7}{|c|}{lirk.: listrank, STRM.: STREAM, ScPC: ScaleParC.} \\
    \hline
  \end{tabular}
  \caption{Workloads Characteristics.}
  \label{Workloads}
\end{table}

To collect granularity message for each memory request, we implement a 3-level cache simulator as a Pin-tool. The detail configuration is listed in table \ref{SystemConfiguration}. We start the cache simulator after each application enters into a representative region. After warm-up the cache simulator with 100 million memory requests, we collect memory traces with granularity and cache access type message. For PARSEC benchmark, we naturally choose the ROI (Region-of-Interest) codes as the region for PARSEC benchmarks; and for all the other benchmarks, we manually skip the initialization phase (such as graph-generation in BFS) and collect memory traces after meaningful work.

\begin{table}[h!]
  \centering
  \begin{tabular}{|c|c|}
    \hline
    \multirow{3}{*}{\textbf{Reorder Buffer}} & 2.7GHz, 256-entry, \\ & max fetch/retire per cycle: 4/2, \\ & 5 pipeline (latency of non-mem instr) \\
    \hline
    \multirow{2}{*}{\textbf{L1 Cache}} & Private, 32KB, 4-way, 64B cache line, \\ & 9 CPU cycles hit (4+5) \\
    \hline
    \multirow{2}{*}{\textbf{L2 Cache}} & Private, 256KB, 8-way, 64B cache line, \\ & 15 CPU cycles hit (10+5) \\
    \hline
    \multirow{2}{*}{\textbf{L3 Cache}} & Shared, 16-way, 64B cache line, \\ & 1MB/core, 45 CPU cycles hit (40+5) \\
    \hline
    \textbf{Memory} & 2 buffer schedulers/MC, \\
    \textbf{Controller} & Read/Write Queue: 64/64 \\
    \hline
    \multirow{2}{*}{\textbf{Link Bus}} & 2.7GHz, point-to-point, \\ & read/write bus width: 16/16 \\
    \hline
    \textbf{Buffer} & FRFCFS \cite{FRFCFSScheduling}, closed page \\
    \textbf{Scheduler}  & Channel-interleave mapping \\
    \hline
    \multicolumn{2}{|c|}{\textbf{DRAM Parameters}} \\
    \hline
    \multirow{2}{*}{\textbf{Memory}} & 2 64-bit Channels, 2 Ranks/Channel, \\ & 8 devices/Rank, 8 sub-ranks/rank, \\ & x8-width sub-rank, 1 device/sub-rank \\
    \hline
    \multirow{4}{*}{\textbf{DRAM Device}} & DDR3-1333MHz, x8, 8 banks, \\ & 32768 Rows/bank, 1024 Columns/Row \\ & 8 KB Row Buffer per Bank, BL=8, \\ & Time and power parameters from \\ & Micron 2Gb SDRAM \cite{DDR3SDRAM} \\
    \hline
  \end{tabular}
  \caption{System Configurations}
  \label{SystemConfiguration}
\end{table}



\subsection {System configurations}

Table \ref{SystemConfiguration} lists the main parameter settings used in the cycle-detailed simulators. Note, the non-memory instruction latency and cache hit latencies listed here are used as the latency of the instruction need to wait in the ROB (in MMAsim) before it could be committed. For example, a L2 cache hit memory access instruction could be committable only after 15 CPU cycles when it is added in the ROB. The baseline memory system has 2 DDR3-1333MHZ channels with dual ranks with 8 DRAM x8 chips each. Each DRAM chip has 8 banks. We fast forward 64 million memory traces for each core (thread), simulate until all the threads have executed at least 100 million instructions.

To evaluate the MIMS, we use the following memory system configuration:

\begin{itemize}
\item DDR: traditional DDRx (3) memory system with fixed coarse access granularity (cache line: 64B), this is the baseline.
\item BOB: Buffer On Board memory system, fixed coarse access granularity, 1 memory request (read/write) per packet, simple packet format without any extra message.
\item MIMS\_one (MI\_1): Message Interfaced based memory system, adopts sub-rank memory organization to support variable-granularity access, 1 request per packet, contains granularity message in packet.
\item MIMS\_multiple (MI\_mul): Message Interfaced based memory system, supports variable-granularity access, multiple requests in a packet.
\end{itemize}

\begin{bfseries}DRAM and Controller Power\end{bfseries}: we evaluate memory power consumption with DRAMsim2 \cite{DRAMSim2} power calculator, which uses the power model developed by Micron Corporation based on the transitions of each bank. The DRAM power is divided into 4 components: background, refresh, activation/precharge, and burst, where background and refresh power is often concluded as static power, activation/precharge and burst power is concluded as dynamic power. Besides DRAM devices, we also take the memory controller power into consider, for it would contribute a significant amount to overall consumption \cite{MemScalePower} (about 20\%). In BOB and MIMS, the controller power is actually referred to simpler controller and buffer scheduler power respectively. For DDR, we adopt the MC power to 8.5W from \cite{AMDPower}; for BOB and MIMS, we adopt the intermediate controller power to 14W as in \cite{BOBMemory}. The controller idle power is set to 50\% of its peak power.

\section {Experimental Results}

In this section,  We first present the performance and power impacts of MIMS in Section 5.1, and then evaluate the effectiveness of combining big-granularity memory requests. We present the effect of memory addresses compression in section 5.3.

\subsection {Performance and power impacts}

In this section, we present simulation results of 16-core systems on FINE and MID granularity workloads. All the workloads are in multiple-thread mode, with each core running one thread. We use the total number of submitted instructions as the metric of performance.

\begin{figure}
  \centering
  \includegraphics[width=0.5\textwidth, height=1.8in]{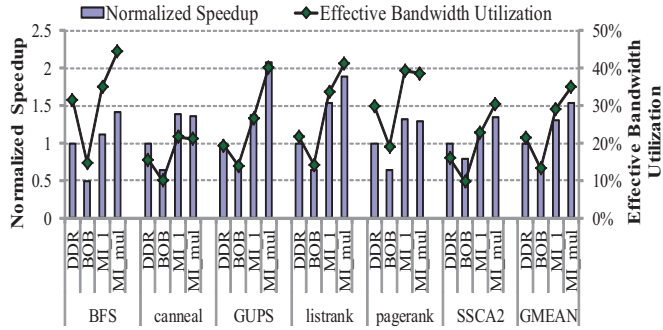}
  \caption{Normalized Speedup and Effective Bandwidth Utilization of different memory system in 16-core configuration, the baseline is DDR}
  \label{Speed_Bandwidth}
\end{figure}

Figure \ref{Speed_Bandwidth} shows the normalized performance speedup and effective bandwidth utilization of different memory system, where the baseline is DDR. For these FINE or MID workloads, such as BFS, canneal, GUPS, fine granularity access would benefit. The BOB performance degrades range from 49.38\% to 78.38\%. This is because that the BOB still uses coarse granularity access, and the intermediate controller would introduce packet overhead and extra latency. On the other side, MIMS\_1 and MIMS\_mul could improve performance because they support variable-granularity, the normalized speedups of MIMS\_1 range from 1.11 to 1.53, and of MIMS\_mul range from 1.29 to 2.08, this indicates that integrating multiple requests in a packet could reduce packet head overheard, thus improve memory performance. The effective bandwidth utilizations nearly have the same variation trend with the speedups in different memory systems. For DDR, they range from 15.58\% to 31.55\%; for BOB, the effective bandwidth utilization is decreased, since each memory request would introduce a packet overhead, along with the wasting bandwidth for transferring useless data in a cache line. The MIMS\_1 could eliminate wasting data but still suffer significant packet overhead. The MIMS\_mul could achieve the best efficiency bandwidth utilization, ranging from 21.15\% to 44.49\%.

\begin{figure}
  \centering
  \includegraphics[width=0.5\textwidth, height=1.8in]{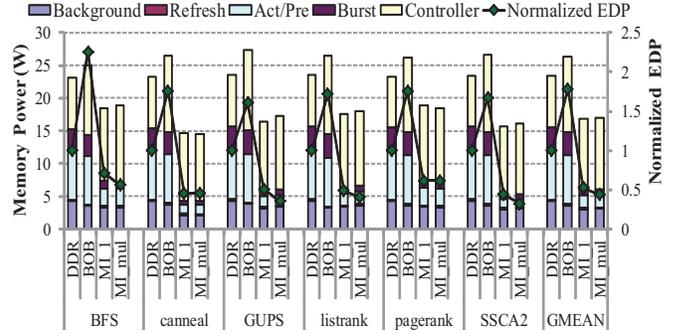}
  \caption{Memory power breakdown and the normalized EDP (Energy*Delay Product, lower is better)}
  \label{Memory_Power_EDP}
\end{figure}

Figure \ref{Memory_Power_EDP} shows the memory power breakdown and the normalized EDP in different memory systems. Here we also consider the power of controller. The average total power for DDR is about 23.38W, and the BOB has a little more power (26.36W), since the intermediate simple controller consumes more power than the on chip MC, the DRAM power of them are nearly the same. The MIMS\_1 and MIMS\_mul could effectively reduce the Activation/Precharge power because each (fine) request only activate/prechare a sub-rank (one DRAM device in our work) with smaller row, and reduce the Burst power because it only read/write the really useful part of data in a cache line (such as 8B data in 64B cache line). Thus the power of MIMS\_1 and MIMS\_mul reduced to 16.90W and 17.13W respectively. The normalized EDP (Energy Delay Product) of BOB reduces about 1.78, this is mainly because the introducing latency. MIMS\_one and MIMS\_multiple could improve EDP by 0.53 and 0.44 respectively, this is because sub-rank could improve memory parallelism and thus reduce queuing delays.

\begin{bfseries}Memory Latency:\end{bfseries}In DDR memory system, the memory latency is categorized into two major categories: Queuing and DRAM Core Access Latency. The Queuing Latency represents latency of a memory request waiting to be scheduled in the Transaction Queue, which has been proved to be the main component of memory latency \cite{RethinkDRAMArch}. The DRAM Core Access Latency represents the latency of executing DDR commands of a memory request in DRAM devices. In MIMS memory system, there is an additional Scheduling latency, which represents the extra processing latency induced by Buffer Scheduler, it includes the SerDes latency, scheduling latency and packet encoding/decoding latency. The Queuing Latency contains both in memory controller (waiting to be packed) and in buffer scheduler (waiting to be issued to DRAM devices) in MIMS.

\begin{figure}
  \centering
  \includegraphics[width=0.5\textwidth, height=1.8in]{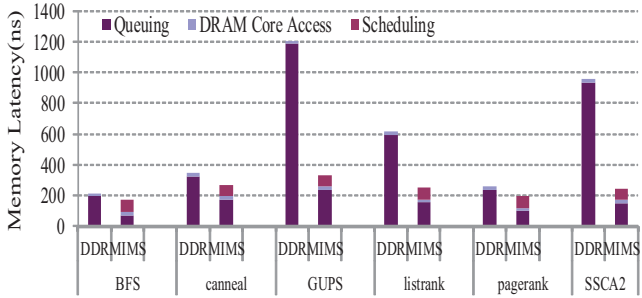}
  \caption{Memory latency breakdown in 16-core configuration.}
  \label{Latency_ns}
\end{figure}

Figure \ref{Latency_ns} shows the memory latency breakdown in 16-core configuration. We can see that for these memory intensive workloads, the Queuing Latency dominates the memory latency, especially for GUPS and SSCA2 application, which could achieve about 1185.67 ns and 933.0 ns respectively in DDR memory system, meanwhile the DRAM Core Access Latency is only 22.22 ns. The reason for it is that these two applications suffer high MPKI as shown in table \ref{Workloads} and the traditional DDR memory system is failed to serve them due to its limited MLP. However, the Queuing Latency could reduce significantly in MIMS, for instance, it reduce to 234.81 ns for GUPS and 147.41 ns for SSCA2, that is because the MIMS adopted sub-rank and it could provide more MLP since each narrow aub-rank could be accessed independently. Even though the intermediate buffer scheduler would induce extra Scheduling Latency, the whole memory latency is reduced for all workloads.

%

\begin{figure}
  \centering
  \includegraphics[width=0.5\textwidth, height=1.8in]{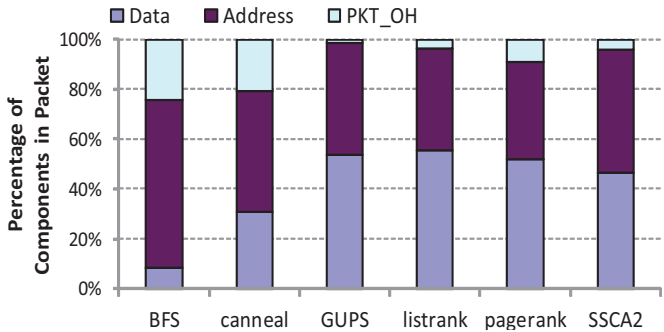}
  \caption{The percentage of data in both Read and Write Packet in MIMS with multiple requests.}
  \label{Percentage_Packet}
\end{figure}

Figure \ref{Percentage_Packet} shows the percentage of different components in packets in MIMS\_multiple memory system. Here we only show the packets in downside (from cpu) bus. For a read packet, it only contains packet overhead (PKT\_OH) and address; for a write packet, it contains data also. We can see that, the address contributes a large portion of packet, it ranges from 39.15\% to 67.39\%, and Data contributes range from 8.24\% to 55.26\%. The BFS benchmark contributes the max address portion, that is because it has a read/write ratio of 9.16 (please refer to Table \ref{Workloads}). The packet overhead has a relatively small portion, because each packet allows to integrate multiple requests, and more requests in a packet, less of the packet overhead. For example, there are about 31.51 requests in each packet, thus the packet overhead is only 1.42\%; but in BFS, there are only about 3.57 requests in each packet, result in about 24.38\%. Observation that address contributes a large portion in a packet gives a good reason for address compression.

\begin{figure}
  \centering
  \includegraphics[width=0.5\textwidth, height=1.8in]{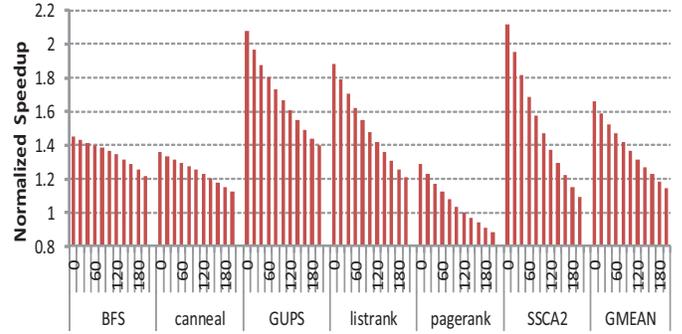}
  \caption{The Normalized Speedup of MIMS as the introduced latency varied from 0 to 200 CPU cycles.}
  \label{Latency_Speedup}
\end{figure}

\begin{bfseries}Latency proportionality of the buffer scheduler in MIMS\end{bfseries}. Buffer scheduler would introduce extra latency for memory requests, since the memory controller needs firstly send requests to buffer scheduler, including: packing multiple requests in the memory controller, SerDes transition, packet transferring on the link bus and packet decoding. Due to different implementation and craft, the introduced latency of buffer scheduler might have proportionality possibilities. In this section, we varied the latency from 0 (perfect) to 200 CPU cycles with a step of 20 CPU cycles to study how the introduced latency would affect the overall memory system performance. Figure \ref{Latency_Speedup} shows the results.

We can see that the latency proportionality of buffer scheduler has a significant impact on the MIMS performance, and the impact is different for different applications. The SSCA2 has the largest slowdown at about 1.07 every 20 more CPU cycles, and the normalized speedup (based on DDR) reduce from 2.12 (with 0) to 1.09 (with 200 cycles). and pagerank even reduced to 0.89, worse than the DDR. All applications still have good speedup with 100 cycls delay.

\subsection {Large Granularity Access}

\begin{figure}
  \centering
  \includegraphics[width=0.5\textwidth, height=1.8in]{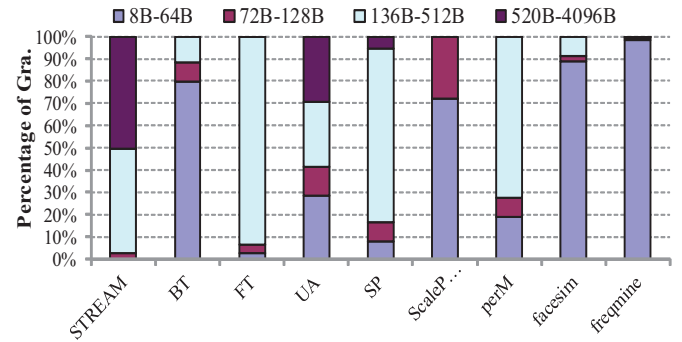}
  \caption{The percentage of combined big granularity memory requests.}
  \label{Percentage_Big_Granularity}
\end{figure}

Besides fine granularity memory access, the MIMS could also support big granularity memory access. For COR applications, continuous memory addresses could be merged into a big access. To simulate a program that can send the trunk information to memory controller, we preprocess memory access traces and merge the memory access traces that access continuous memory address into one request within a instruction-window. We set the preprocess-window size to 256, which means we can combined 256 traces once, but the merged trace should not be larger than 4KB for read, and 512B for write. Figure \ref{Percentage_Big_Granularity} shows the size of combined request in our experiments.

\begin{figure}
  \centering
  \includegraphics[width=0.5\textwidth, height=1.8in]{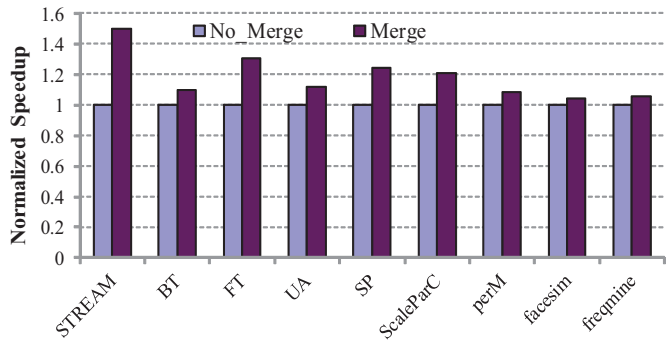}
  \caption{Normalized Speedup with combining big granularity memory requests.}
  \label{Speedup_Big_Granularity}
\end{figure}

 Figure \ref{Speedup_Big_Granularity} shows the speed up result. It can be seen that MIMS can gain better performance through large granularity. Compared to the original memory access, the bigger proportion of large granularity is, the better performance MIMS can achieve. STREAM has 50\% improvement. To get the large granularity information from application, software hint or compiler support will be better than hardware detection.

\subsection {Address Compression}

In this section, we evaluate the address compression in packet. We will show that even with some simple compression algorithms, it would achieve substantial compression ratio. This is due to the natural locality among adjacent addresses, especially for coarse granularity workloads. 3 compression algorithms are evaluated:
\begin{itemize}
\item Single\_Base: the first address in a packet is chosen as the single base, all the other address would be represented as the difference to the base (delta value). The base address needs to be placed in the packet.
\item Multi\_Base\_Inline: the Packet Generator and Packet Decoder synchronously maintain a base address table which contains multiple base addresses. Each address is represented as an index and a difference if it could matches one of the base. otherwise, this address would serve as a base address, and the LRU old entry in the table is selected to evict. This new base address needs to be sent to the Packet Decoder, and make the two table keep synchronous.
\item Multi\_Base\_Offline: extended from the multi\_base\_inline algorithm. At each packet, each base would automatically updated by the last address that could be compressed. This simple learning strategy would keep base address in step and increase the possibility to hit with compression.
\end{itemize}

\begin{figure}
  \centering
  \includegraphics[width=0.5\textwidth, height=1.8in]{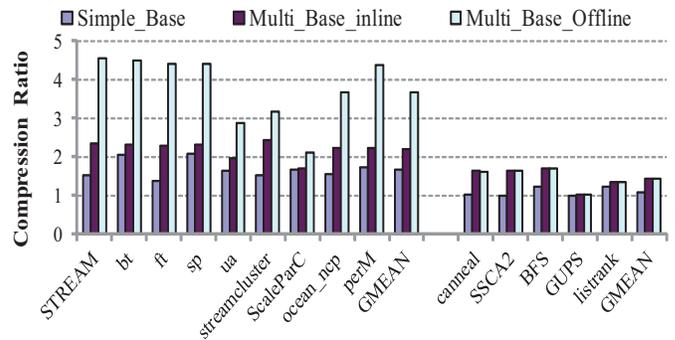}
  \caption{Address Compression Ratio.}
  \label{CompressionRatio}
\end{figure}

In our experiments, for coarse granularity applications, the number of base address is 8 (requires 3 bits index), and the bits of difference is set to 8; and for fine granularity applications, the number of base address is 8, and the bits of difference is 24. Figure \ref{CompressionRatio} shows the compression ratio with the above three compression algorithm. For coarse granularity applications (left part), the Single\_Base could achieve about 1.66, the Multi\_Base\_Inline could achieve about 2.19, and the Multi\_Base\_Offline could achieve about 3.68, where the STREAM application could get the highest compression ratio of 4.56. These indicate that there exists many opportunity to exploit address compression. However, for fine granularity applications, since the memory access pattern is random, the compression ratio is relatively low, the Single\_Base is only 1.08, and the Multi\_Base\_Inline and the Multi\_Base\_Offline nearly achieve the same compression ratio, about 1.44.

\section {Conclusions and Future Work}

In this paper, a Message Interface based Memory System (MIMS) is proposed. By decoupling memory access with memory organization and associating semantic information with memory request, MIMS provides new opportunities to solve existing memory problems. Experimental results show that MIMS is able to improve parallelism and bandwidth utilization for fine granularity applications, to keep locality for trunk memory access and enable effective inter-requests address compression.

Using message interface instead of traditional bus might open a new road for memory system design. In the futre we will extend MIMS to support more operations and investigate on implementation issues.



\bstctlcite{bstctl:etal, bstctl:nodash, bstctl:simpurl}
\bibliographystyle{IEEEtranS}
\bibliography{references}

\end{document}